\documentclass[preprint2]{aastex}
\usepackage{graphicx}
\usepackage{txfonts}
\usepackage{natbib}
\usepackage[scaled]{helvet}
\usepackage{epsfig}
\usepackage{url}
\bibpunct{(}{)}{;}{a}{}{,}
\usepackage[normalem]{ulem}
\usepackage{color}

\begin{document}
\title{V409 Tau As Another AA Tau: Photometric Observations of Stellar Occultations by the Circumstellar Disk}
\author{Joseph E. Rodriguez$^{1}$, Joshua Pepper$^{2,1}$, Keivan G. Stassun$^{1,3}$, Robert J. Siverd$^{4,1}$, Phillip Cargile$^{5,1}$, David A. Weintraub$^1$, Thomas G. Beatty$^{6,7}$, B. Scott Gaudi$^8$, Eric E. Mamajek$^9$, Nicole Sanchez$^{3,1}$}

\affil{$^1$Department of Physics and Astronomy, Vanderbilt University, 6301 Stevenson Center, Nashville, TN 37235, USA}
\affil{$^2$Department of Physics, Lehigh University, 16 Memorial Drive East, Bethlehem, PA 18015, USA}
\affil{$^3$Department of Physics, Fisk University, 1000 17th Avenue North, Nashville, TN 37208, USA}
\affil{$^4$ Las Cumbres Observatory Global Telescope Network, 6740 Cortona Dr., Suite 102, Santa Barbara, CA 93117, USA}
\affil{$^5$ Harvard-Smithsonian Center for Astrophysics, 60 Garden St, Cambridge, MA 02138, USA}
\affil{$^6$Department of Astronomy \& Astrophysics, The Pennsylvania State University, 525 Davey Lab, University Park, PA 16802}
\affil{$^7$Center for Exoplanets and Habitable Worlds, The Pennsylvania State University, 525 Davey Lab, University Park, PA 16802}
\affil{$^8$Department of Astronomy, The Ohio State University, Columbus, OH 43210, USA}
\affil{$^9$Department of Physics and Astronomy, University of Rochester, Rochester, NY 14627-0171, USA}

\shorttitle{V409 Tau}

\begin{abstract}
AA Tau is a well studied young stellar object that presents many of the photometric characteristics of a Classical T Tauri star (CTTS), including short-timescale stochastic variability attributed to spots and/or accretion as well as long-duration dimming events attributed to occultations by vertical features (e.g., warps) in its circumstellar disk. We present new photometric observations of AA Tau from the Kilodegree Extremely Little Telescope North (KELT-North)
which reveal a deep, extended dimming event in 2011, which we show supports the interpretation by \citet{Bouvier:2013} of an occultation by a high-density feature in the circumstellar disk located $>$8 AU from the star.
We also present KELT-North observations of V409 Tau, a relatively unstudied young stellar object also in Taurus-Auriga,
showing short timescale erratic variability, along with two separate long and deep dimming events, one from January 2009 through late October 2010, and the other from March 2012 until at least September 2013. We interpret both dimming events to have lasted more than 600 days, each with a depth of $\sim$1.4 mag. From a spectral energy distribution analysis, we propose that V409 Tau is most likely surrounded by a circumstellar disk viewed nearly edge-on, and using Keplerian timescale arguments we interpret the deep dimmings of V409 Tau as occultations from one or more features within this disk $\gtrsim$10 AU from the star. In both AA Tau and V409 Tau, the usual CTTS short-timescale variations associated with accretion processes close to the stars continue during the occultations, further supporting the distant occulting material interpretation. Like AA Tau, V409 Tau serves as a laboratory for studying the detailed structure of the protoplanetary environments of T~Tauri disks, specifically disk structures that may be signposts of planet formation at many AU out in the disk. We also provide a table of all currently known disk-occulting young stars as a convenient reference for future work on such objects.

\end{abstract}

\maketitle
\section{\bf{Introduction}}
The circumstellar environments of young stellar objects (YSOs) involve complex dynamical interactions between dust and gas that directly influence the formation of planets. UX Orionis stars (UXors), which are one specific category of YSOs, are a class of pre-main sequence stars that have circumstellar disks and experience large aperiodic photometric dimming events. These events range in depth and duration but can be up to 3 magnitudes in the $V$-band on timescales of days to months. Most UXor stars exhibit infrared excesses in their spectral energy distributions (SEDs) that are generally interpreted as being due to emission from circumstellar disks. Some UXor stars show minima lasting months to years with long-term (decade) changes in median base line brightness \citep{Rostopchina:1999}. During the minima, a color reversal from red to blue is normally observed together with an increase in polarization. The initial dimming and reddening of light is thought to originate from high column densities of dust, while the bluing during minima and the polarization are caused by an increase in the scattered light \citep{Bibo:1991, Grinin:1991, Grinin:1998, Waters:1998}. 

Hydrodynamical fluctuations in the stars' circumtellar disks have been suggested to explain the large dimming events seen in UXor stars \citep{Dullemond:2003}. Occultations of the host star by dust clumps in their circumstellar disk has also been proposed as another explanation. \citep{Wenzel:1969, Grinin:1988, Voshchinnikov:1989, Grinin:1998, Grady:2000}. If the dimmings are caused by features in a geometrically thin disk, then the disk would need to be seen very nearly edge-on \citep{Grinin:1991, Grinin:1996, Herbst:1999, Bertout:2000}. 
However, the UXor variations described above typically occur in optically visible pre-main sequence stars that are surrounded by circumstellar disks viewed at less edge-on inclinations of $45^{\circ}$ - $68^{\circ}$, indicating the disks must be flared and/or possess warps or large-scale features with large vertical scale heights \citep{Natta:2000}. Even though UXor stars are usually early type Herbig Ae and Be stars, a few are Classical T Tauri Stars (CTTS) that display similar photometric variability, such as the late K star AA Tau \citep{Bouvier:1999}. It has been found that about 10$\%$ of late type CTTS display the UXor photometric dimming variability, perhaps reflecting the required geometric and line-of-sight orientation \citep{Bertout:2000}. A list of CTTS that are candidate UXors is provided in Table 1.

\begin{table}[ht]
\caption{CTTS that are candidate UXor stars (G or later)}
\begin{tabular}{ | c | c | c | c |}
\hline
Target  &   Spectral Type   &    Reference \\
\hline
CO Ori  &   G5  &  \citet{Eaton:1995} \\
RY Tau  &   F8-K1   &   \citet{Eaton:1995}\\
RY Lupi &   G0  &   \citet{Eaton:1995}\\
DK Tau  &   K6  &  \citet{Oudmaijer:2001}\\
CB 34V  &  G5 & \citet{Tackett:2003}\\
SU Aur  &   G2  &   \citet{Unruh:2004}\\
UY Aur  &  G5 &  \citet{Berdnikov:2010}\\
AA Tau  & K7 &  \citet{Bouvier:2013} \\
V409 Tau   & M1.5    &   This work\\
\hline
\end{tabular}
\footnotesize{{\bf Notes.} The spectral types listed here are from SIMBAD except for AA Tau and V409, the references for which are provided in \S2.}
\end{table}

One way to better understand the structure and evolution of circumstellar disks is to observe a star being occulted by circumstellar material. AA Tau is one of the most best-studied stellar systems displaying both short and long term photometric variability, a common characteristic of T Tauri stars \citep{Vrba:1993, Bouvier:1999, Bouvier:2003, Bouvier:2007}. Based on long-term monitoring, AA Tau had remained at a constant brightness from 1978 until 2011 \citep{Grankin:2007, Bouvier:2013}. In 2011, AA Tau dimmed by $\sim$2 mag \citep{Bouvier:2013} and has not returned to its normal brightness as of UT 2013 December 9 (The end of our data set). \citet{Bouvier:2013} concluded that the dimming is likely the result of a density increase in the occulting region of the circumstellar disk that is located at least 8 AU from the host star, assuming the occulter is in a Keplerian orbit. Based on the 2 years of constant brightness, \citet{Bouvier:2013} conclude that the azimuthal extent of the occulter is at least 30$^{\circ}$. During this dimming, the system experiences a color reversal from red to blue, a known characteristic of UXor stars \citep{Bouvier:2013}. 

Another example of a young star being occulted by circumstellar material is the well-studied system RW Aurigae. In late 2010, the known T Tauri system RW Aurigae, which contains at least two stellar components (RW Aurigae A and B), experienced a $\sim$2 magnitude dimming event for $\sim$180 days \citep{Rodriguez:2013}. The RW Aurigae system has been photometrically observed since the late 1890's \citep{Beck:2001}, but no event of similar depth and duration had ever been seen prior to 2010. Detailed investigations of the system suggest that there was a recent close fly-by of RW Aur A by RW Aur B, tidally disrupting the outer portions of the circumstellar disk of RW Aur A \citep{Cabrit:2006}. The dimming was attributed to a portion of the tidally disrupted circumstellar disk occulting the primary component, RW Aur A. Subsequent modeling of the star-disk encounter by \cite{Dai:2015} supports this interpretation. This system shows that binary star interactions can sculpt circumstellar disks. There are other systems in the literature that display both periodic and non-periodic large dimming events, which as of yet are unexplained, that may also be caused by circumstellar material \citep{Carroll:1991, Kearns:1998, Chiang:2004, Winn:2004, Bouvier:2007,Grinin:2008, Plavchan:2008, Kloppenborg:2010, Mamajek:2012, Rattenbury:2014}.



Objects like AA Tau and RW Aur, that display large photometric dimming events caused by their circumstellar disks, give us opportunities for studying the evolution of the circumstellar environments of young stars and perhaps even embryonic planets in disks. In this paper, we present new photometric observations of the 2011 dimming of AA Tau. The observations of the sudden dimming of AA Tau support the previous interpretation by \citet{Bouvier:2013} that the dimming is caused by a region of enhanced density in the circumstellar disk. We also present new photometric observations of V409 Tau that show that it underwent two separate dimming events. In the observations of V409 Tau we present below, we see two clear dimming events that are separated by a short period of time and are similar to the dimming event of AA Tau. Our observations of V409 Tau prior to the two dimming events are much sparser in cadence but appear consistent with the typical short-term variability associated with late type T Tauri stars. We interpret the dimming of V409 Tau, analogously to the dimming event of AA Tau, as likely due to photometric variability resulting from density variations in the disk of V409 Tau.\footnote{Throughout this paper, we refer to the reductions of brightness of V409 as ``dimmings", and not ``eclipses".  Although we propose that the observed dimmings are caused by eclipses of the star by intervening material, we avoid the term ``eclipse" when referring to the observations to maintain generality in the descriptions of the data.} These long dimming events are known characteristics of UXor stars, and thus we can add V409 Tau to the short list of late-type stars showing this behavior. 

The paper is organized as follows. We introduce the known characteristics of the V409 Tau system in \S2, illustrating the complex stellar environment. In \S3, we describe the photometric observations, and then discuss the photometric properties of the data in \S4. In \S5, we present several interpretations of the light curve and discuss their plausibility. We summarize our results and conclusions in \S6. 

\section{\bf Known Characteristics}
In this section, we present a review of the known physical and observational parameters of the stars V409 Tau and AA Tau.


\subsection{\bf V409 Tau}
V409 Tau ($\alpha$ = 04h 18m 10.785s, $\delta$ = $+25^{\circ}$ 19$\arcmin$ 57.39$\arcsec$; V$\sim$13.34 \citep{Zacharias:2012}) was determined by \citet{Kenyon:1994} to be a member of the Taurus-Auriga association using IRAS observations. Spectroscopic observations by \citet{Luhman:2009} show V409 Tau to be a M1.5 star, with $T_{\rm eff}$ = 3632 K, and provide strong evidence confirming it as a member of the Taurus-Auriga association. \citet{Gorynja:1968} first classified V409 Tau as a variable star.  Using the Trans-atlantic Exoplanet Survey (TrES), \citet{Xiao:2012} measured periodic photometric variability of 4.754 days with an amplitude of 0.3 mag. \citet{Andrews:2013} estimated the age of V409 Tau to be variously 1.86 Myr, 5.012 Myr, and 3.47 Myr using three separate pre-main sequence stellar models \citep{D'Antona:1997, Baraffe:1998, Siess:2000}. From these models, the stellar mass was estimated to be $\sim$0.40 $M_{\sun}$, $\sim$0.63 $M_{\sun}$ and $\sim$0.43 $M_{\sun}$ respectively. 

\subsection{\bf AA Tau}
AA Tau is a low-mass, photometrically variable, system in the Taurus-Auriga association that has been previously classified as a UXor star. It has a spectral type of $\sim$K7, V$\sim$12.2, $T_{\rm eff}$ = 4030K, radius of 1.85 $R_{\sun}$ and a mass of 0.85 $M_{\sun}$ \citep{Bouvier:1999}. Both photometric and spectroscopic monitoring provide evidence that the system is experiencing magnetospheric accretion \citep{Bouvier:2003}. The outer portion of the AA Tau disk has an inclination of $\sim71^{\circ}$ while the inner part of the disk is warped and misaligned with respect to the outer disk \citep{Cox:2013}. \citet{Bouvier:1999} observed a $\sim$8.5 days, $\sim$1.4 mag quasi-periodic variability in the $UBVRI$ filters where the $B-V$ color only showed a 0.1 mag change on the timescale of weeks. This $\sim$8 day photometric periodicity, first reported by \citet{Vrba:1993}, has been attributed to the magnetically warped inner disk periodically occulting the host star \citep{Bouvier:2007}. These eclipses have presented with varying depths, which provides evidence that the occulter is changing size or opacity on the timescale of days. It is likely that the inner warp and the accretion streams from the disk onto the stellar surface are at similar locations in the inner disk, orbiting with a period of $\sim$8 days. The accretion is likely varying, directly affecting the warp, and likely causing the change in the eclipse depth \citep{Bouvier:2007}. In a study of $>$ 500 variables in the Orion Nebula Cluster, \citet{Rice:2015} identify 73 young stars that show “AA Tau” like short quasi periodic dimming events.

In 2011, AA Tau dimmed by $\sim$2 mag. \citet{Bouvier:2013} claim that the near-IR colors during AA Tau's dimmed state and an observed bluer slope in the red/NIR spectrum from XSHOOTER \citep{D'Odorico:2006, Vernet:2011} suggest that the system is experiencing 3-4 mags of visual extinction compared to its normal bright state. \citet{Bouvier:2013} argue this because they required a much higher extinction (A$_v$ $>$ 3) to match the red and near-IR wavelength regimes in the XSHOOTER spectrum. They argue that the reason the system only shows $\sim$2 mag depth event in the V band (and not a 3-4 mag event) is that the system experiences a color reversal becoming bluer at optical wavelengths during the dimming. This color reversal is a common feature of UXor stars.

Importantly, while the $\sim$8 day quasi-periodic variability was not observed during the initial stage of the 2011 dimming event \citep{Bouvier:2013}, it was seen during a later stage of the dimming event in late 2012/2013, for a 3 month period by \cite{Bouvier:2013}. This indicates that the source of the long timescale dimming is likely situated outside the inner disk region that is the likely source of the short-timescale accretion variability. Indeed, on the basis of Keplerian arguments, \citet{Bouvier:2013} suggested that the obscuring material must be situated at $\sim$8 AU from the star.

\begin{figure*}[!ht]
\centering\epsfig{file=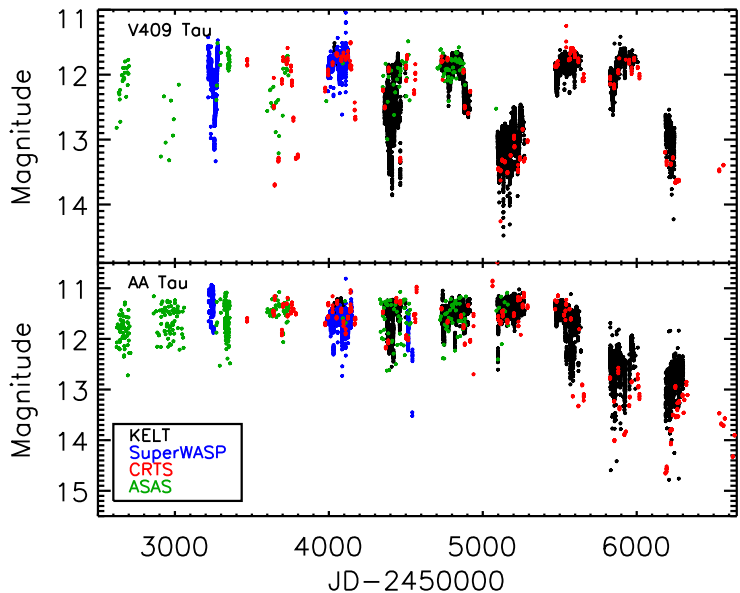,clip=,width=0.6\linewidth}
\caption{KELT-North (Black), SuperWASP (Blue), CRTS (Red) and the ASAS (Green) light curves of V409 Tau (Top) and AA Tau (Bottom) from 2004 to 2013. A vertical offset has been applied to the KELT, SuperWASP and ASAS data to match pre-dimming magnitudes of AA Tau to the $V$ band observation by CRTS. The same vertical offset has been applied to the V409 Tau observations. Only the CRTS data are in $V$-band magnitudes whereas the other observations are in very broad band magnitudes that we approximate to $V$-band but no attempt has been made to place all the data on the same absolute scale.}
 \label{fig_full_LC}
\end{figure*}

\section{\bf Photometric Observations}
V409 Tau and AA Tau have been observed in several photometric surveys over the past two decades. We present new photometric data for both targets in Figure~\ref{fig_full_LC}.

\subsection{Archival Data}
The All Sky Automated Survey (ASAS) is a photometric survey with the goal of observing as much of the southern sky as possible in order to study any and all kinds of photometric variability in the $I$-band. \citet{Pojamanski:1997} describes the data acquisition and reduction techniques. ASAS observed V409 Tau from UT 2002 December 13 to UT 2009 September 12, obtaining 153 observations. ASAS also observed AA Tau from  UT 2002 December 13 to UT 2009 November 30, collecting 376  observations.

The Catalina Real-time Transient Survey (CRTS) is a wide-field photometric survey designed to detect variable sources on the time scale of minutes to years using a $V$-band filter. The data used in the work described here were from Catalina Survey Data Release 2\footnote{http://nesssi.cacr.caltech.edu/DataRelease/}. More information on the CRTS observations and data reduction can be found in \citet{Drake:2009}. CRTS observed V409 Tau from UT 2005 April 9 to UT 2013 September 27, acquiring 350 observations. CRTS observed AA Tau from UT 2005 April 8 to UT 2013 December 9, obtaining 429 observations.

The Wide Angle Search for Planets (SuperWASP) is a photometric survey for transiting extrasolar planets with a cadence of a few minutes in broad filter centered on 550nm. SuperWASP observed V409 Tau, first in 2004 from July 29 to September 30 and then again from UT 2006 September 17 to UT 2007 January 24, for a total of 3695 images of V409 Tau. SuperWASP observed AA Tau in 3 separate seasons: UT 2004 August 2 to UT 2004 September 4, UT 2006 September 11 to UT 2007 February 15 and UT 2008 February 18 to UT 2008 March 17, acquiring 2744 observations. The SuperWASP public archive is described in \citet{Butters:2010}.

Using the 40/50/100 cm Schmidt telescope at Asiago, \citet{Romano:1975} obtained 24 non-filtered observations of V409 Tau from UT 1962 October 22, until UT 1964 July 13. The observations showed a brightening from 1962 to 1964. This data is not publicly available and therefore we do not include it in our light curve analysis. We refer back to these observations when interpreting our results to understand the V409 Tau system in \S5.2.

\subsection{KELT-North}

The Kilodegree Extremely Little Telescope (KELT-North) project is an ongoing, wide-field ($26^{\circ}$ $\times$ $26^{\circ}$) survey for transiting planets around bright stars ($V$ = 8-10). The survey uses two telescopes, KELT-South (Sutherland, South Africa) and KELT-North (Sonita, Arizona), and observe in a broad $R$-band filter, with a $\sim$15 minute cadence and a typical photometric precision for stars of $V\sim11$ of $\sim0.04$ mag \citep{Pepper:2007, Pepper:2012}. The KELT data are reduced using a heavily modified version of the ISIS software package \citep{Alard:1998, Alard:2000}, described further in \S2 of \citet{Siverd:2012}\footnote{Much of the reduction software is publicly available:  http://verdis.phy.vanderbilt.edu}. V409 Tau and AA Tau are both located in KELT-North Field 03, which is centered on $\alpha$ = 3h 58m 12s, $\delta$ = $59^{\circ}$ 32$\arcmin$ 24$\arcsec$. For the work presented in this paper, all of the KELT observations come from the KELT-North telescope. KELT-North observed this field for 7 seasons from UT 2006 October 26 to UT 2013 January 9, obtaining $\sim$9100 images. All data shown has a relative photometric error less than 20\% RMS. Table 2 shows the start and end date for each KELT-North season. A portion of the KELT-North photometric data set is shown in table 3.
\begin{table}[ht]
\caption{KELT-North Table of Observing Seasons}
\begin{tabular}{ | c | c | c | }
\hline
Season & UT Start Date & UT End Date\\
\hline
1 & 2006 October 26 & 2007 January 17\\
2 & 2007 September 19 & 2008 February 2\\
3 & 2008 September 24 & 2009 March 26\\
4 & 2009 September 22 & 2010 March 16\\
5 & 2010 October 2 & 2011 March 17\\
6 & 2011 September 22 & 2012 March 21\\
7 & 2012 September 17 & 2013 January 9th\\
\hline
\end{tabular}
\end{table}

\begin{table*}[ht]
\centering
\caption{KELT-North photometric observations of V409 Tau and AA Tau}
\begin{tabular}{ | c | c | c | c | c | }
\hline
JD$_{\! {\rm  TT}}$   &         V409 Tau    &    V409 Tau  &         AA Tau     &    AA Tau \\
  &         Relative Mag\tablenotemark{a}    &     Photometric Errors\tablenotemark{b} &          Relative Mag\tablenotemark{a}    &     Photometric Errors\tablenotemark{b}\\
\hline
2454034.731241 & -0.181 & 0.054 & 0.084 & 0.044\\
2454034.735862 & -0.209 & 0.053 & 0.026 & 0.041\\
2454034.740481 & -0.205 & 0.046 & 0.109 & 0.044\\
2454034.745101 & -0.195 & 0.052 & 0.081 & 0.044\\
2454034.758962 & -0.209 & 0.049 & 0.138 & 0.064\\
2454034.763581 & -0.239 & 0.047 & 0.030 & 0.039\\
2454034.768202 & -0.205 & 0.048 & 0.082 & 0.043\\
2454034.772822 & -0.228 & 0.044 & 0.096 & 0.0439\\
2454034.791297 & -0.235 & 0.047 & 0.094 & 0.044\\
2454034.795917 & -0.158 & 0.051 &  0.135 & 0.044\\
\hline
\end{tabular}

\footnotesize{{\bf Notes.} The data shown in Table 3 is published in its entirety in the electronic edition of this paper for both V409 Tau and AA Tau. 

$^{\rm a}$Relative KELT-North Instrumental magnitude. The median of the KELT-North Instrumental magnitude has been subtracted off.  

$^{\rm b}$Photometric errors for instrumental KELT-North magnitudes. True per-point magnitude errors must fold in 0.036 mag systematic errors.}
\end{table*}

\subsection{CARMA 3mm}
We observed the V409 Tau system using the Combined Array for Research in Millimeter Astronomy (CARMA) for $\sim$3.5 hours in the 3mm continuum. Specifically, we used the CARMA 15 E configuration consisting of nine 6.1m and six 10.4m antennas. All observations were taken on UT 2014 August 8. Along with observing V409 Tau, we observed Uranus, 3C84 and 0510+180 as the flux, passband and gain calibrators, respectively. The observations and flux value presented in this paper were reduced and measured using the MIRIAD software program, described in detail by \citet{Sault:1995}. 

\section{Analysis and Results}
Here we present and discuss the photometric properties of AA Tau and V409 Tau (data shown in Figure \ref{fig_full_LC}). Our analysis focuses on the SuperWASP, ASAS, CRTS and KELT-North photometric data. 

\subsection{\bf AA Tau}

AA Tau has been classified previously as a UXor star, displaying both short term quasi-periodic variability and non-periodic large dimming events. The KELT-North photometric data confirm known characteristics of the AA Tau system presented in the literature.  

\begin{figure}[!ht]
\centering\epsfig{file=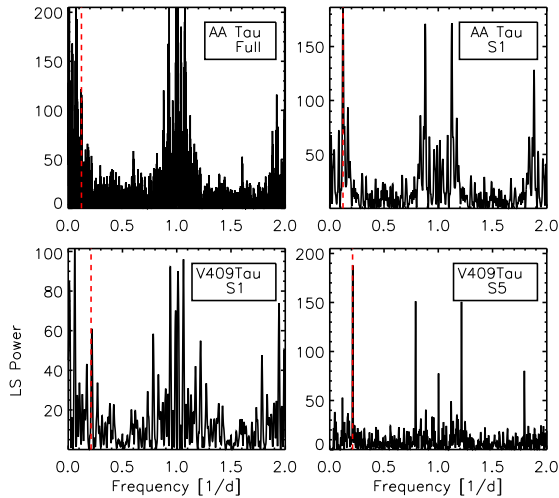,clip=,width=0.97\linewidth}
\caption{LS periodicity analysis of the KELT-North photometric data. Top Row: LS periodogram of for the Full KELT-North AA Tau data set (Left) and KELT-North season 1 (right). The vertical red dashed line corresponds to the 8.2 day period found by \citet{Vrba:1993}. Bottom Row: KELT-North season 1 V409 Tau data set (Left) and KELT-North season 5 (Right). The vertical red dashed line is the 4.574 day period found by \citet{Xiao:2012}. The large peak at 1.0 and 2.0 cycles/day in all periodograms is a diurnal alias of the long period variability.}

\label{fig_LS}
\end{figure}

\begin{figure}[!ht]
\centering\epsfig{file=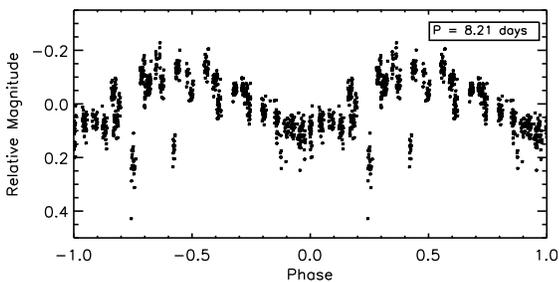,clip=,width=0.97\linewidth}
\caption{KELT-North Season 1 light curve of AA Tau phased to an 8.21 day period recovered from LS analysis.}
\label{fig_phase}
\end{figure}

First, after a long quiescent period, AA Tau decreased in brightness by $\sim$2 mag in 2011 \citep{Grankin:2007, Bouvier:2013}.  Second, the pre-dimming obervations of AA Tau in all data sets show a $\sim$1.0 mag amplitude variability that is common to most classical T Tauri stars, as first shown by \cite{Herbst:1994}. YSOs will display both periodic and non-periodic variability, typically on the timescales of days to weeks. We use the Lomb-Scargle (LS) period search in the VARTOOLS analysis package to search for periodic variability \citep{Hartman:2012}. The Lomb-Scargle (LS) periodicity analysis is designed to search for small sinusoidal periodic signals in unevenly sampled time-series data \citep{Lomb:1976,Scargle:1982,Press:1989}. The goal of our LS analysis was to recover the previously known $\sim$8 day period for AA Tau. Our LS analysis of the first KELT-North season of AA Tau shows a periodicity of $\sim$8.2 days (Figure \ref{fig_LS}). The phased light curve of AA Tau at a period of 8.21 days is shown in Figure \ref{fig_phase}. This is similar to the periodicity found by \citet{Vrba:1993, Bouvier:1999} and is interpreted as the magnetically warped inner disk occulting the host star at a semi-major axis of a few stellar radii. The large peaks at 1.0 and 2.0 cycles/day in all periodograms is a diurnal alias of the long period variability. Finally, the pre-dimming observations of AA Tau in all data sets show variability of up to 1.0 mag amplitude that is common characteristic of most classical T Tauri stars, as first shown by \citet{Herbst:1994}. 


\begin{figure*}[!ht]
\centering\epsfig{file=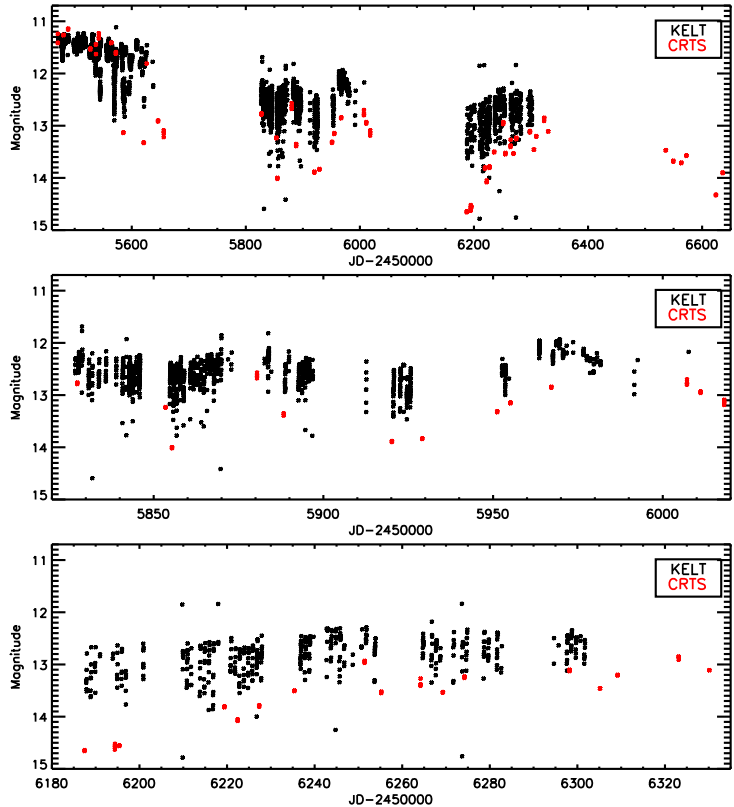,clip=,width=0.6\linewidth}
\caption{(Top) KELT-North (Black) and CRTS (Red) light curves of AA Tau during the 2011 sudden dimming event. (Middle) A zoom-in of the first full KELT-North season during the dimming. (Bottom) A zoom-in of the second full KELT-North season during the dimming.}
 \label{fig_AA_Tau_zoom}
\end{figure*}

We confirm the previous dimming event in KELT-North survey data. The KELT-North and CRTS observations of AA Tau show that the dimming event began in late November of 2010 and has remained consistently fainter through the end of our photometric observations, December 2013 (see Figure \ref{fig_AA_Tau_zoom}). This means the sudden dimming event has lasted $\sim$3 years and is likely still occurring.

During the dimming, we do not recover the $\sim$8 day periodicity. During the dimming event, AA Tau is near the sensitivity limit of KELT-North. Therefore, we cannot conclusively rule out the possibility of a periodicity during the dimming. However, the $\sim$1 mag variability that is characteristic of T Tauri stars, is clearly observed in both the KELT-North and CRTS data sets. Moreover, \citet{Bouvier:2013} did clearly observe the $\sim$8 day quasi-periodic variability during a 3 month period in 2012/2013. 
They suggest the reason that the 8-day periodicity is not observed throughout the 2011 dimming is that the observed light has become dominated by scattered light originating from high latitudes on the star. Consequently, the 8-day modulation of the equatorial latitudes by the magnetically warped inner disk seen during AA Tau bright state is not continuously observed, even though the 8-day periodicity does in fact continue \citep{Bouvier:2013}.

\subsection{\bf V409 Tau}
In the KELT-North and CRTS data, we observe two separate instances where V409 Tau has clearly dimmed below the known brightness level for an extended period of time. In this section, we discuss the observational characteristics of these events. We also analyze the general photometric variability seen both during and outside the two prominent dimming events.

\subsubsection{Out-of-Dimming Variability}
\begin{figure}[!ht]
\centering\epsfig{file=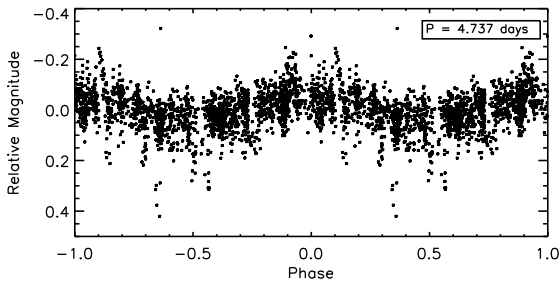,clip=,width=0.97\linewidth}
\caption{KELT-North Season 5 light curve of V409 Tau phased to a 4.723 day period recovered from our LS analysis. A 5 harmonic Fourier series was fit to the data to remove long term trends. }
\label{fig_phase_V409}
\end{figure}

Young stars tend to display both periodic and aperiodic photometric variability, both of which can be caused by circumstellar extinction and accretion of material onto the star's surface \citep{Herbst:1994, Grinin:2004, Petrov:2007}. \citet{Xiao:2012} found a 4.754 day, 0.3 mag amplitude periodicity in TrES photometric observations of V409 Tau \citep{Alonso:2007}.  The goal of our LS analysis was to recover the previously known period for V409 Tau. Using LS analysis to search for periodicities of less than or equal to 1000 days, we do not find any evidence of a significant periodicity in the combined KELT, SuperWASP, CRTS, and ASAS photometric data sets. However, using only the KELT-North season 1 data we recover a 4.5 day period and using only KELT-North season 5, we recover a 4.7 day period (see Figure \ref{fig_LS}). We fit a Fourier series, with 5 harmonics, to remove long term trends in season 5 of KELT. A phased plot (4.737 day period) of the KELT-North season 5 data (with the long term trends removed) is shown in Figure \ref{fig_phase_V409}. These periods are similar to the period seen by \citet{Xiao:2012}.

Prior to the dimming events seen in 2009 and 2012, V409 Tau appears to display some photometric variability with amplitude varying from 0.5-2 mags. This variability is also seen during both dimming events. Unfortunately, we have not found any archival photometric observations of V409 Tau prior to 2002, which hinders our ability to determine the longevity of the variability we present here. There are reported observations from UT 1962 October 22 until UT 1964 July 13 by \citet{Romano:1975}. These observations display a light curve that begins at a magnitude of $\sim$15 and ends at $\sim$13.5. Interestingly, this feature has similar properties to the dimming events seen by the KELT-North and CRTS data.

The V409 Tau light curves do show dimming features of much shallower depths and shorter duration in the KELT, SuperWASP, CRTS, and ASAS from 2002 to 2007. The data sampling is much sparser during this period, which hinders our ability to determine if these are true variations or if they are just artifacts of the data quality and sampling. Between the 2010 and 2013 events, for about $\sim$500 days, V409 Tau brightened back to its normal $V$ mag of $\sim$12. During this period, the typical chaotic variability is clearly observed. 

\subsubsection{ 2009-2010 Dimming Event }
In late January 2009, V409 Tau dimmed significantly from a pre-dimming brightness of $V\sim$11.8 down to $V\sim$13.2 (Figure \ref{fig_dim_LC}a). In the KELT-North and CRTS data, this dimming begins at a JD$\sim$2454850 (Terrestrial Time (TT)). It is clear, however, that this dimming event has a peak depth below the long-term median of $\sim$1.4 mag and that it sustained for over a year. Since there is no known companion to V409 Tau, we assume the entire dimming is caused by a decrease in the brightness of the host star, corresponding to a decrease in flux of 72\%.

After the $\sim$200 day long observing gap beginning at JD$\sim$2454900 (TT), V409 Tau is clearly fainter by $\sim$1.4 mag and stays fainter through the extent of the observing season (until JD$\sim$2455300 (TT)). Over the course of this observing season (KELT-North observing season 4), where V409 Tau is in a dim state, the brightness of the system increases by a few tenths of a magnitude. After the following seasonal observing gap, V409 Tau is back at its nominal pre-dimming brightness level of $V$$\sim$11.7, the median brightness of KELT-North observing season 3. Due to the end of the seasonal observing gap, we cannot determine exactly when the dimming event of V409 Tau ended and can only place an upper limit on the duration of the event.

If the event began at JD$\sim$2454850 (TT), then the maximum duration we estimate is $\sim$630 days (the red vertical lines in Figure \ref{fig_dim_LC}), however this is necessarily an upper limit on the dimming duration. We estimate the ingress of this event to be $\sim$240 days (grey shaded region in Figure \ref{fig_dim_LC}) but this is not well constrained due to the seasonal observing gaps, and could have been much shorter.

A significant amount of structure can be seen during the dimming. Unfortunately, during the dimming event, the star's brightness is quite low, nearly consistent with zero flux for KELT-North, thus complicating the analysis. The CRTS observations appear to show a slightly deeper event then seen in KELT-North. This is likely attributed to the CRTS data being in the $V$ filter while the KELT-North observations are in a broad $R$ filter. However, there is a short-term variability in the KELT-North and CRTS data starting around JD$\sim$2455170 (TT) and ending at JD$\sim$2455289 (TT), and that is similar in amplitude to that seen outside of the dimming. This variability may continue for a longer but the KELT-North and CRTS data sets ended due to the seasonal observing window. In any case, there is evidence for continuing variability during the dimming, as in AA Tau, and that is similar in nature to the short timescale variations seen outside of the dimming event.

\subsubsection{2012-2013 Dimming Event}
\begin{figure*}[!ht]
\centering\epsfig{file=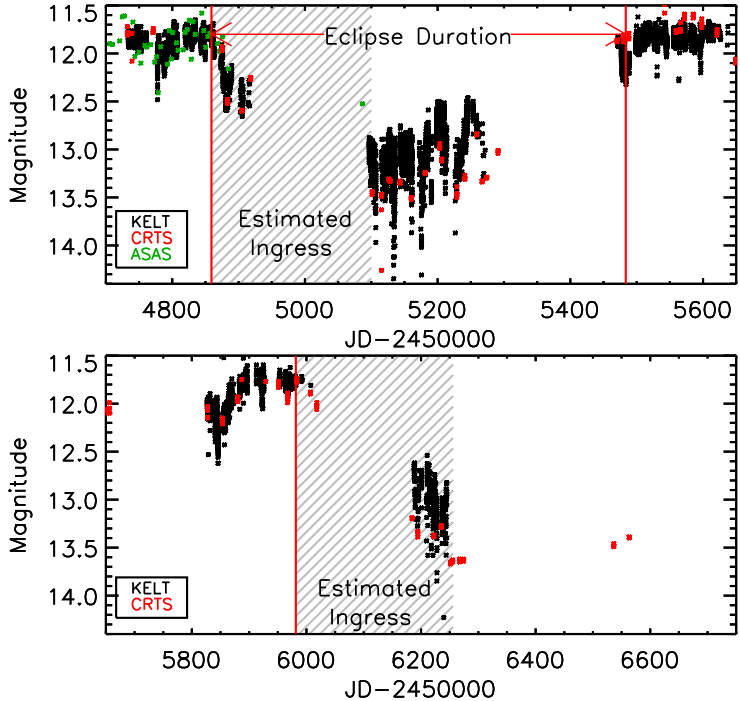,clip=,width=0.6\linewidth}
\caption{KELT-North (Black), CRTS (red) and the ASAS (green) light curves of V409 Tau during the first (top) and second (bottom) dimming events. The shaded region corresponds to the estimated duration of the events ingress.}
 \label{fig_dim_LC}
\end{figure*}

In May 2012, V409 Tau experienced another large dimming event, from a pre-dimming brightness of $V\sim$11.8 down to $V\sim$13.2 (Figure \ref{fig_dim_LC}b). This event is quite similar to the 2010 event discussed in the previous subsection. We estimate the ingress of this event to be $\sim$275 days, slightly longer than the 2009 event, but again this parameter is not well constrained. This event lasted through the extent of the KELT-North and CRTS data sets. Similar to the 2009-2010 dimming, the CRTS observations appear to show a slightly deeper event then seen in KELT-North. For this dimming event, both the ingress and potential egress appear to be located in the seasonal observing gaps for V409 Tau, hindering our ability to determine key characteristics such as the duration ingress and egress. Thus, we cannot determine whether this dimming event was of a similar duration to the 2010 event, which were separated by $\sim$1000 days. We can rule out the possibility that this specific $\sim$600 day long, 1.4 mag depth dimming event is periodic because we should have detected another event around JD$\sim$2454200(TT). Although there is a gap in the combined photometric coverage from all data sets around this time, it is not large enough to completely miss another event similar in duration to what was seen in 2010 and possibly 2012. However, a $\sim$1.5 mag brightening event was observed from 1962 till 1964 by \citet{Romano:1975} using observations from the 40/50/100 cm Schmidt telescope at Asiago. It is possible that \citet{Romano:1975} measured the recovery of another dimming even and this type of dimming is periodic in V409 Tau. However, with the lack of photometric coverage between the early 1960's and the start of our photometric data set (2002), it difficult to be conclusive. 

\subsubsection{Spectral Energy Distribution Analysis}

We assembled the available broadband flux measurements of V409 Tau (not observed during one of the large dimming events) from the literature and also measured a new flux at 3 mm using the CARMA array (see Table \ref{tab:seddata}). We fit these fluxes using the star+disk spectral energy distribution (SED) model grid of 
\citet{Robitaille2006,Robitaille2007}, based on the radiation transfer code of 
\citet{Whitney2003a,Whitney2003b}. Briefly, these SED models represent young stars with disks and/or envelopes for physical parameters spanning a large range of stellar masses, temperatures, and luminosities, and for disk parameters spanning a large range of sizes, structures, accretion rates, and inclinations on the plane of the sky. Additional free parameters include the line-of-sight extinction and distance to the system.

\begin{table}[ht]
\caption{Archival flux measurements of V409 Tau used in the SED analysis.\label{tab:seddata}}
\begin{tabular}{ | c | c | c | c | }
\hline
Band & Flux\tablenotemark{a} & Error\tablenotemark{b} & Reference\\
\hline
NUV & 0.030 & 0.021 & GALEX \\
$u'$ & 17.5 & 1.0 & SDSS \\
$g'$ & 15.5 & 0.7 & SDSS \\
$r'$ & 13.0 & 0.7 & SDSS \\
$i'$ & 12.0 & 0.7 & SDSS \\
$z$ & 11.7 & 0.7 & SDSS \\
$J$ & 10.7 & 0.5 & 2MASS \\
$H$ & 9.6 & 0.5 & 2MASS \\
$K_S$ & 9.0 & 0.5 & 2MASS \\
WISE1 & 8.3 & 0.3 & WISE \\
WISE2 & 8.0 & 0.3 & WISE \\
WISE3 & 5.6 & 0.3 & WISE \\
WISE4 & 3.8 & 0.3 & WISE \\
IRAC1 & 8.1 & 0.3 & Spitzer \\
IRAC2 & 7.8 & 0.3 & Spitzer \\
IRAC3 & 7.3 & 0.3 & Spitzer \\
IRAC4 & 6.3 & 0.3 & Spitzer \\
MIPS1 & 4.4 & 0.3 & Spitzer \\
MIPS2 & 1.7 & 0.3 & Spitzer \\
MIPS3 & $-$1.45 & Upper limit & Spitzer \\
0.89 mm & 48.8 mJy & 22.2 mJy & \citet{Luhman:2009}\\
1.3 mm & 18.7 mJy & 1.4 mJy & \citet{Luhman:2009}\\
3 mm & 2.87 mJy & 0.28 mJy & this work \\
\hline
\end{tabular}
\tablenotetext{a}{Magnitudes unless otherwise indicated.}
\tablenotetext{b}{Single-epoch errors have been inflated to reflect time variability of the source.}
\end{table}

\begin{figure}[!ht]
\centering\epsfig{file=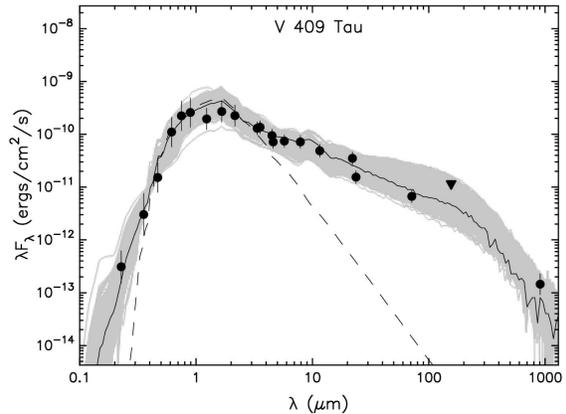,clip=,width=0.99\linewidth}
\caption{Spectral Energy Distribution fit for the V409 Tau system. Symbols with error bars represent flux measurements with uncertainties, and inverted triangles represent 3$\sigma$ upper limits (see Table~\ref{tab:seddata}). The dashed curve represents the photosphere while the gray curves represent all of the star+disk models that are consistent with the data  to application of the $T_{\rm eff}$, $A_J$, and distance criteria. The solid curve represents the final best-fit SED including all observational constraints (see the text).
\label{SED_Fit}}
\end{figure}

Figure~\ref{SED_Fit} shows the observed SED and the family of best-fit models satisfying a goodness-of-fit criterion of $\Delta\chi^2 / \chi_{\rm min}^2 < 3$ per data point 
\cite[see, e.g.,][for discussion of this statistical criterion]{Robitaille2007}. Note that the SED models included in the \citet{Robitaille2006} grid extend only to $\lambda = 1$mm so Figure~\ref{SED_Fit} extends only to 1~mm. The SED models in gray represent the various star+disk parameter combinations that are formally consistent with the data according to the above $\chi^2$ criteria. 

To further narrow the range of possible SED models, we additionally required the model SED parameters to include a stellar $T_{\rm eff}$ within 300~K of that determined by \citet{Luhman:2009}, a total line-of-sight extinction within 0.3 mag of the $A_J = 1.3$ also determined by \citet{Luhman:2009}, and a distance of 140$\pm$40 pc (i.e., within 40 pc of the nominal Taurus-Auriga association distance). Applying these additional constraints significantly reduces the SED models that can fit the data to within the $\chi^2$ goodness-of-fit criterion above, leading to a single model in the \citet{Robitaille2006} grid (solid black SED in Figure \ref{SED_Fit}). This model has a disk that is inclined at an angle of 81$^\circ$, i.e., nearly edge-on. We take this as suggestive that the V409 Tau disk is fully consistent with being nearly, but not precisely, edge-on while satisfying the other observed constraints for the system. The new CARMA observation at 3~mm is not included in the fit but is consistent with a simple extrapolation of the best-fit model.

\section{\bf Interpretation}
In this section we present the most plausible interpretation of both the sudden dimming event of AA Tau and the dimming events seen from V409 Tau. For AA Tau, we concur with the previous observations and interpretations by \citet{Bouvier:2013} that the dimming is likely caused by an occultation of the host star by a higher density region of the circumstellar disk. For V409 Tau, we also interpret the dimming events to be caused by an occultation of the host star by a one or more features in the circumstellar disk, predicted from our SED analysis. Below we explore the involvement of each system's circumstellar disk in the large dimming events observed.  

\subsection{\bf AA Tau}
\citet{Bouvier:2013} suggests that the deep and sudden dimming of AA Tau is likely due to an increase in extinction from an overdense region in the known circumstellar disk on a Keplerian timescale. In this section, we calculate the parameters of this feature from the KELT-North and CRTS data, assuming the feature is located in the disk plane. We compare our results with the work done by \citet{Bouvier:2013}.

Using observations from the Crimean Astrophysical Observatory (CrAO), \citet{Bouvier:2013} determined that AA Tau took $\sim$200 days to reach minimum brightness. From the KELT-North and CRTS data, we visually estimate the ingress of the AA Tau dimming to be $\sim$300 days long, beginning in late November or early December of 2010. Since AA Tau was in a consistently bright state for 24 years and the host star is $\sim$0.8$M_{\sun}$, \citet{Bouvier:2013} determined that the occulter would have a semi-major axis $\ge$ 7.7 AU. Given the known inclination of the circumstellar disk around AA Tau, we can estimate the height of the feature that caused the dimming, assuming it is located in the disk plane. Using a semi-major axis of 7.7 AU and a disk inclination of $\sim71^{\circ}$, this would require a scale height of $\sim$7.3 AU for the occulter to cross the line of sight. The duration of the event is $\sim$800 days and it has not begun to recover to its pre-dimming median brightness through the extent of our data set. Using a stellar mass of 0.85$M_{\sun}$ and a semi-major axis of 7.7 AU, this would require the azimuthal extent of the feature to be over $46^{\circ}$, or $>6$ AU.

Given the measured and calculated parameters of the occulting feature, we can model the dimming event as an occultation of the host star by a large body in a Keplerian orbit. This body cannot have a leading edge perpendicular to its direction of motion because it would be moving too slowly to be located within the body of the circumstellar disk. Rather, we model the occulting object to have a leading edge that is slightly inclined from its direction of motion, i.e., ``wedge" shaped. Assuming the feature causing the dimming is located within the estimated radius of the AA Tau disk, i.e., within 215 AU \citep{Kitamura:2002}, and assuming an ingress timescale of 200 days, this would require the leading edge of the feature to be $<$4.6$^{\circ}$ ($<$3.2$^{\circ}$ for ingress of 300 days), close to parallel with its direction of motion. Using an ingress duration of 200 days and a wedge angle of 5$^{\circ}$, the transverse velocity of the occulter would be $2(1.84 R_{\sun})/(T_{ingress} \sin\theta) \sim$1.7 km~s$^{-1}$. Assuming Keplerian motion, this would place the occulting object near the edge of the circumstellar disk. For the body to be located at 7.7 AU, the minimum semi-major axis determined by \citet{Bouvier:2013}, the leading edge of the occulter would need to have a wedge angle of $<$1$^{\circ}$. The long duration of the AA Tau dimming can be interpreted in multiple ways, including a wedge, a fuzzy edge, and/or scattered light. Since we can't definitively determine which of these is correct, we adopt the wedge model presented above as a way of developing an illustrative model that is representative of the timescales and spatial scales involved.

\subsection{\bf V409 Tau}
From the SED analysis, we have shown that V409 Tau has an infrared excess indicative of a circumstellar disk, specifically one that is nearly edge-on. We interpret the dimming events as occultations of the host star by one or more features in the circumstellar disk. This interpretation is supported by our observations since the $\sim$1 magnitude variability, seen prior to the dimming events, is apparent during the dimming. We believe the short-term variability arises on or near the stellar surface. Therefore, an occultation by material much farther away in the disk would have no effect on the fractional amplitude of the short term variability.

\begin{figure}[!ht]
\centering\epsfig{file=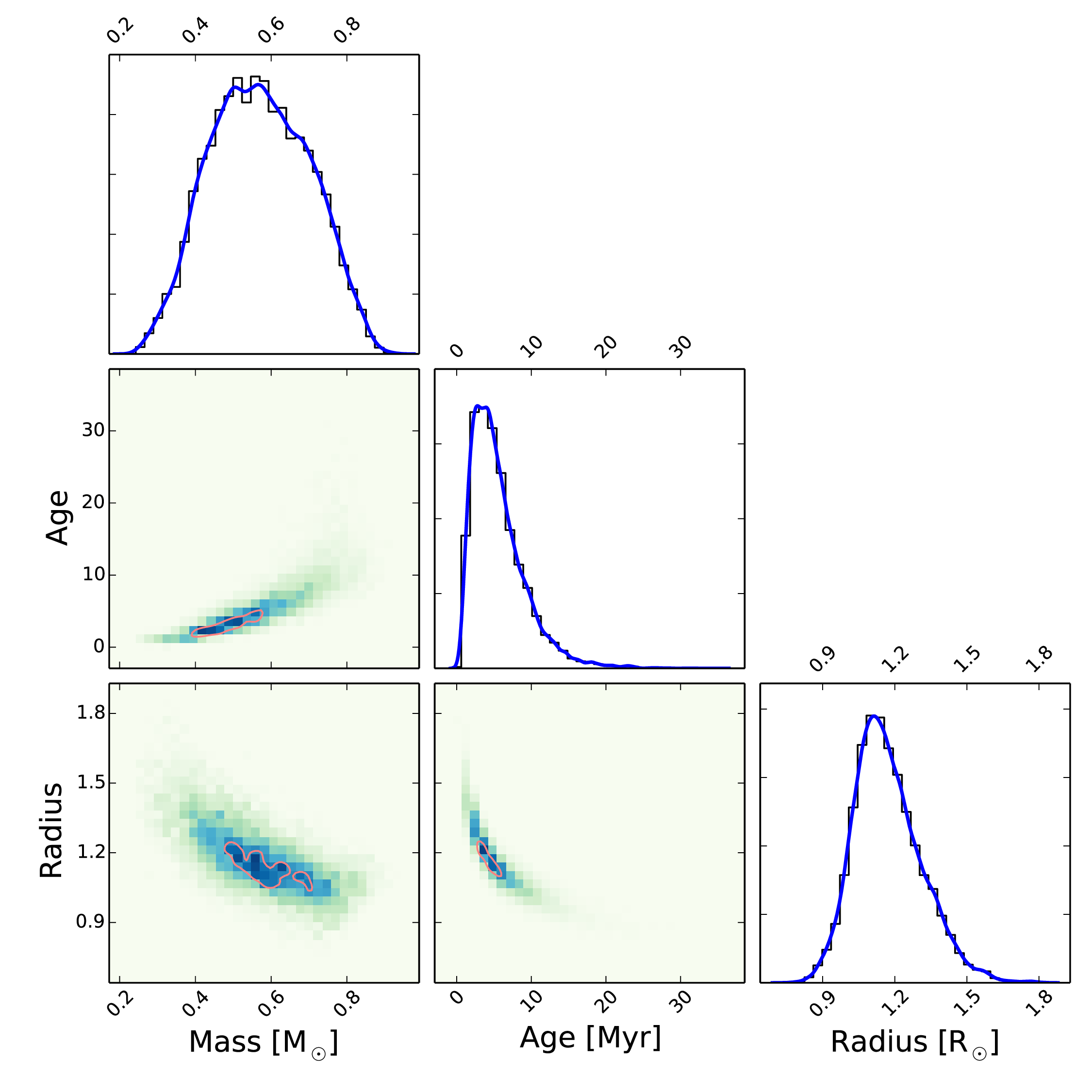,clip=,width=0.96\linewidth}
\caption{The posterior distributions of mass, age, and radius, and our determined parameters (w/ 68\% confidence interval uncertainties)}
\label{Radius_MCMC}
\end{figure}

Through simple kinematic and geometric arguments, we are able to estimate certain parameters of the occulting body. One key parameter necessary for calculating the properties of the occulting body is the radius of V409 Tau. Using prior observables of Log(T$_{eff}$) = 3.56 $\pm$ 0.02, Log(L/L$_{\sun}$) = -0.59$\pm$ 0.08 and [Fe/H] = -0.01 $\pm$ 0.05 for V409 Tau \citep{Andrews:2013, D'Orazi:2011}, we use the Dartmouth Stellar Evolution Program’s young stellar models within a Markov Chain Monte Carlo (MCMC) to derive posterior distributions for the mass, age and radius of V409 Tau (see Figure \ref{Radius_MCMC}) and the most probable parameters (with 68\% confidence interval uncertainties) \citep{Dotter:2008}. From this analysis, we determine an age of $2.78_{-0.42}^{+6.10}$ Myr, a mass of $0.57_{-0.13}^{+0.16}$ M$_{\sun}$ and a radius of $1.11_{-0.07}^{+0.20}$ R$_{\sun}$.

We can repeat the same kind of model calculations for  V409 Tau as we did for AA Tau. That is an occultation of the host star by a large body in a Keplerian orbit, possessing a slightly inclined wedge-shaped leading edge. Beginning with the 2009 dimming, we estimate the ingress timescale to be $\sim$240 days. Since we do not have an estimate of the outer disk radius for V409 Tau, we constrain the feature to the known radius of the AA Tau circumstellar disk, $\sim$215 AU \citep{Kitamura:2002}. At 215 AU, this would require the leading edge of the occulter to have a wedge angle $<$ 2.8$^{\circ}$ for an ingress of 240 days (see \S4.2.2). We estimate the ingress of the 2012 event to be close to 275 days (see \S4.2.3) which constrains the leading edge wedge angle to be $<$ 2.5$^{\circ}$. Both of these estimated leading-edge wedge angles correspond to the feature being located at the outer edge of the circumstellar disk. 

Observations by \citet{Romano:1975} show a brightening of V409 Tau from a visual magnitude of 15 to $\sim$13.5 over the time-span of 630 days. From their data, it is possible that they observed the end of another dimming event. If they did observe another dimming event, then the full duration would likely be longer than the 2009 and 2012 events seen in the KELT-North and CRTS data. Without other photometric observations between our data and the observations in the 1960's by \citet{Romano:1975}, we are unable to better constrain a potential periodicity. There is just over 46 years between the end of the 1962 brightening event and the end of the 2009 event. Since there are two consecutive potential occultations of V409 Tau in our data and one from the 1960's data by \citet{Romano:1975}, if we assume they are all related, we can interpret them in two ways: 1) They are caused by two separate features orbiting in similar locations of the circumstellar disk. 2) A single feature is responsible for all three events.

\begin{figure}[!ht]
\centering\epsfig{file=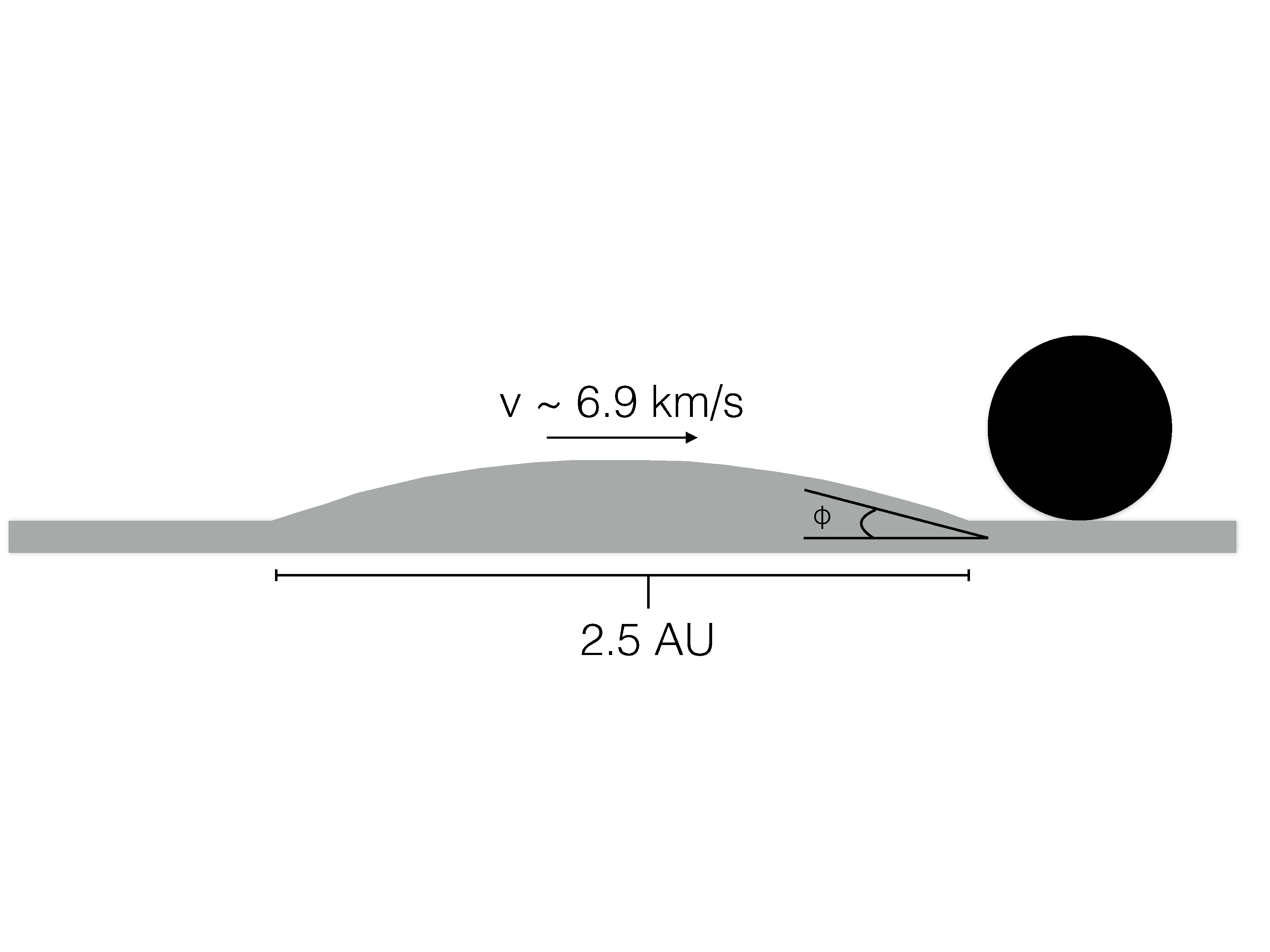,clip=, width=0.99\linewidth}
\caption{A diagram displaying the theoretical warp or ``wedge-shaped" feature in the V409 Tau disk that has a shallow leading edge or ``wedge angle". Not to scale. $\Phi$ corresponds to the leading edge angle or wedge angle.}
\label{V409_drawing}
\end{figure}

If the 2009 and 2012 events are from two separate features within the circumstellar disk with a minimum period of $\sim$46.25 years (the time between the 1962 and 2009 events), assuming Keplerian motion and using our derived parameters for the stellar mass of V409 Tau, this would require the semi-major axis of the occulting feature to be $>$10.7 AU. At this distance from the star, the feature would have a very shallow ``wedge-shaped" leading-edge angle of $\sim$0.6$^{\circ}$ corresponding to a transverse velocity of $2(1.11 R_{\sun})/(T_{ingress} \sin\theta) \sim$6.9 km~s$^{-1}$. Using the full estimated maximum duration of the 2009-2010 dimming, $\sim$630 days, this would require the feature to have a width of $\sim$2.5 AU (see Figure \ref{V409_drawing}. The beginning of the 2012 dimming is $\sim$1130 days after the 2009 one. Using the same assumptions as above and that the second feature is orbiting with a period $\sim$1130 days longer than the first, this would require a semi-major axis of $>$11.1 AU, velocity $\sim$6.8 km s$^{-1}$ and a leading edge wedge angle of $\sim$0.5$^{\circ}$. We interpret this feature as two separate warps or perturbations in the circumstellar disk that have crossed our line of sight. Another possibility is that we are seeing two separate dust clumps (or high density regions), located in a similar location of the disk. Simulations show that dust clumps can form in groups and then combine in the disk over time, which could be the beginning processes of planetary formation \citep{Fromang:2005}. 

The second interpretation is that a single feature caused all three dimming events observed. Assuming this scenario would imply a periodicity of $\sim$1130 days, the time between the 2009 and 2012 events. This would place the occulting feature $\sim$1.7 AU from the star. This would also imply that the duration and depth is changing on a dynamical timescale, since we should have observed two more dimming events in 2006 and 2002. Examining the KELT-North observations around those dates, there is a hint of potential dimming events, but the behavior is more consistent with the $\sim$1 mag short period photometric variability that are commonly known for T Tauri stars \citep{Herbst:1994}. However, since the circumstellar disk around V409 Tau is likely close to being edge-on (from our SED analysis), any small fluctuation or change in the thickness of the disk would strongly affect the optical brightness. \citet{Nelson:2000} showed that heating and cooling of the inner disk (inside 10 AU from the host star) would show fluctuations in the spatial distribution of the grains on the timescale $<$ 10 years. Since both observed dimming events have durations estimated to be shorter than 10 years, they are consistent with heating and cooling timescales in the disk. 

From our SED analysis (see \S4.2.4), our best fit model has a disk inclination of $\sim$81$^{\circ}$. Again, this is only suggestive and we only conclude that the disk is likely close to but not exactly edge-on. From our analysis, we estimate that the feature would be $>$10.7 AU from the host star if we interpret the events in our data to be caused by different features in the disk, but related to the event seen in the early 1960’s. If we interpret the events to be caused by one feature, the estimated semi-major axis of the feature would be ~2 AU. Adopting the the disk inclination of 81$^{\circ}$ at a semi-major axis of 10.7 AU and 2 AU, the required height of the warp would only need to be 2 AU and 0.3 AU respectively. Therefore, it would only take a relatively small perturbation to cross our line of sight. Also, we estimate from kinematics the width of the feature to be 2.5 AU from kinematics. We would expect that a warp would be wider than tall which is possible in this scenario. These estimates don’t consider that the disk could flare out as a function of radius, which has been seen in many other disks \citep{Espaillat:2010}.


\section{\bf Conclusions}
The study of stars being occulted by circumstellar material can provide useful information about the environments and dynamics of YSOs. This information may be of great value for our understanding of how planetary systems evolve. New observations of AA Tau from the KELT-North survey allow us to expand on the analysis performed by \citet{Bouvier:2013}. The sudden dimming that began in late 2010 appears to still be occurring without any signs of recovery. We are able to use the data to constrain proprieties of the occulting feature, assuming Keplerian motion. 

New observations of V409 Tau from the KELT-North survey show that the system experienced two separate, $\ge$630 day long $\sim$1.4 mag dimming events in early 2009 and mid-2012. Our observations also show a photometric variability on the timescale of days to weeks occurring before and during the events. Both of these characteristics are a common behavior of UXor stars. The events were confirmed independently by the Catalina Real-time Transient Survey. A brightening of V409 Tau was seen in the early 1960's by \citet{Romano:1975}. It is possible that this event was another dimming followed by a brightening. Our SED analysis models the V409 Tau system to have a nearly edge-on disk. Using KELT-North photometric observations of AA Tau as a direct comparison, we interpret the V409 Tau dimming events as occultations of the host star by one or more features in the nearly edge-on circumstellar disk. We see that both stars display both the short period (days to weeks) variability and long term dimming events (months to years) that are commonly associated with T Tauri and UXor stars. 

The timescales of the two dimming events in V409 Tau are consistent with a fluctuation in the disk's thickness, but it is also possible that they are caused by an occultation of the host star by a warp or perturbation in the circumstellar disk. The short period chaotic variability, presumably due to spots and/or accretion on the stellar surface, is also apparent during the dimming events of V409 Tau, further supporting the occultation scenario by circumstellar disk structures far from the inner disk where the short-timescale variations presumably arise. 

These results motivate additional observations to monitor the V409 Tau system for another dimming event. Like AA Tau, V409 Tau should serve as a valuable laboratory for detailed studies of the structure of protoplanetary environments around low-mass pre-main-sequence stars. The use of photometric surveys like KELT to discover and characterize the occulation of young stars by their circumstellar disks will be applicable to future surveys such as the Large Synoptic Survey Telescope (LSST). With LSST, we will be able to significantly increase the sample size of these rare and valuable systems.

\acknowledgements
We would like to thank Alice Quillen and the entire KELT team for their helpful discussions.

We have used observational data from the ASAS photometric survey and we are thankful for the observations and data reduction performed.

The CSS survey is funded by the National Aeronautics and Space Administration under Grant No. NNG05GF22G issued through the Science Mission Directorate Near-Earth Objects Observations Program. The CRTS survey is supported by the U.S. National Science Foundation under grants AST-0909182 and AST-1313422.

Early work on KELT-North was supported by NASA Grant NNG04GO70G. J.A.P. and K.G.S. acknowledge support from the Vanderbilt Office of the Provost through the Vanderbilt Initiative in Data-intensive Astrophysics.

Work by B.S.G. and T.G.B. was partially supported by NSF CAREER Grant AST-1056524.

This work has made use of NASA's Astrophysics Data System and the SIMBAD database operated at CDS, Strasbourg, France.
EEM acknowledges support from NSF award AST-1313029.

This paper makes use of data from the first public release of the SuperWASP data \citep{Butters:2010} as provided by the SuperWASP consortium and services at the NASA Exoplanet Archive, which is operated by the California Institute of Technology, under contract with the National Aeronautics and Space Administration under the Exoplanet Exploration Program.

Support for CARMA construction was derived from the Gordon and Betty Moore Foundation, the Kenneth T. and Eileen L. Norris Foundation, the James S. McDonnell Foundation, the Associates of the California Institute of Technology, the University of Chicago, the states of California, Illinois, and Maryland, and the National Science Foundation. Ongoing CARMA development and operations are supported by the National Science Foundation under a cooperative agreement, and by the CARMA partner universities.

\bibliographystyle{apj}

\bibliography{V409_Tau}

\end{document}